\documentclass[floatfix, aps, amsmath, nofootinbib, twocolumn, 10pt]{revtex4}

\usepackage{listings}
\usepackage{graphicx}
\usepackage{bm}
\usepackage{rotating}
\usepackage{array}
\usepackage{amsmath}
\usepackage{amssymb} %花体字母加粗
\usepackage{mathrsfs} %花体字母
\usepackage{cancel}
\usepackage{subfig}
\usepackage{float}
\usepackage{caption}

\def\({\left(}
\def\){\right)}
\def\[{\left[}
\def\]{\right]}

\def\e{\begin{equation}}
\def\q{\end{equation}}
\def\m{\begin{eqnarray}}
\def\n{\end{eqnarray}}

\begin{document}
%\title{Frozen States of Neutron Stars with  Magnetic Monopoles}
\title{Frozen Neutron Stars}% Force line breaks with \\
\author{Chen Tan$^{1, 2}$}
\author{Yong-Qiang Wang$^{1, 2}$}
\thanks{Corresponding author: {yqwang@lzu.edu.cn}}
\affiliation{$^{1}$Lanzhou Center for Theoretical Physics, Key Laboratory of Theoretical Physics of Gansu Province, 
	School of Physical Science and Technology, Lanzhou University, Lanzhou 730000, China}
\affiliation{$^{2}$Institute of Theoretical Physics $\&$ Research Center of Gravitation, Lanzhou University, Lanzhou 730000, China}

\date{\today}% It is always \today, today, 

\begin{abstract}
We investigate neutron stars with nonlinear magnetic monopoles in the framework of the Einstein-nonlinear electrodynamics model, specifically within the Bardeen and Hayward models. Solving the modified Tolman-Oppenheimer-Volkoff equations for three different equations of state, we find that upon reaching the critical magnetic charge $q_{c}$, neutron stars enter frozen states characterized by the critical horizon. This extends the concept of frozen states to compact objects composed of ordinary matter (non-field matter), thereby offering a new perspective for related research. 
\end{abstract} 

\maketitle

%\tableofcontents

\section{Introduction}
\label{sec:intro}
Neutron stars (NSs), the ultracompact remnants of core-collapse supernovae, provide a unique environment for probing fundamental physics under extreme conditions—including supranuclear densities \cite{Oertel:2016bki}, strong gravity \cite{LIGOScientific:2017zic,Shao:2022koz}, and intense magnetic fields \cite{Duncan:1992hi, Turolla:2015mwa}. For example, magnetic fields in magnetars are thought to exceed $10^{14}-10^{15}[\rm{G}]$ , profoundly influencing their thermal \cite{Vigano:2013lea} and rotational evolution \cite{Thompson:1996pe}, outburst activities \cite{Thompson:1995gw}, and gravitational wave emission \cite{DallOsso:2008kll}. The strong gravitational and magnetic fields of neutron stars make them efficient natural traps for magnetic monopoles—hypothetical particles predicted by grand unification theories (GUTs) \cite{tHooft:1974kcl}. Although no free magnetic monopoles have been observed to date, they are predicted in several GUTs scenarios and could be effectively captured and accumulated in neutron stars \cite{Bonnardeau:1979cr,Harvey:1982py}. There, they catalyze proton decay \cite{Callan:1982ac, Rubakov:1981rg}, profoundly affecting stellar activity and evolution. 

When focusing on the gravitational effects of magnetic monopoles rather than their nuclear physical effects, we find that traditional treatments of magnetic monopoles, rooted in linear Maxwell electrodynamics, are inadequate for describing their gracitational fields within compact objects. Such a classical framework inevitably leads to divergent electromagnetic energy densities. When coupled to gravity via the Einstein field equations, this divergence results in unavoidable spacetime singularities. This limitation motivates the use of nonlinear electrodynamics (NED), where self-interactions regularize the monopole solution. Well-known geometric models such as the Bardeen and Hayward spacetimes \cite{Bardeen:1968, Hayward:2005gi} (originally proposed as regular black hole solutions) can be effectively interpreted as the gravitational field of nonlinear magnetic monopoles within specific NED frameworks \cite{Hayward:2005gi, Ayon-Beato:1998hmi, Ayon-Beato:2000mjt}. These developments have stimulated broader interest in and discussion of nonlinear magnetic monopoles. 

Recent studies have incorporated the Bardeen and Hayward models into investigations of boson stars \cite{Wang:2023tdz,Yue:2023sep,Zhao:2025hdg,Chicaiza-Medina:2025wul}, leading to the construction of boson star solutions with nonlinear magnetic monopoles. 
Interestingly, these studies reveal that no black hole solution with an event horizon (EH) exists within this framework. When the magnetic charge exceeds a critical value, a frozen boson star incorporating a nonlinear magnetic monopole is realized in the limit $\omega\to0$. In this regime, the scalar field converges inside the critical horizon and decays rapidly beyond it. Within the star’s critical horizon, the metric component $-g_{tt}$ approaches zero. For a distant observer, such stellar configurations may exhibit properties analogous to those of an extremal black hole. The characteristics of these solutions are consistent with those of a frozen star, which is a theoretical model first arising from Oppenheimer and Snyder's analysis of gravitational collapse in black hole formation  \cite{Oppenheimer:1939ue}, and later formally named by Y. Zel'dovich and I. Novikov  \cite{zeldovichbookorpaper}. When observed from a distant perspective, the collapse of an ultra-compact object appears
to occur over an extended period, creating that the star is frozen at their own gravitational radius \cite{Ruffini:1971bza}. 

In this paper, we investigate neutron stars that incorporate nonlinear magnetic monopoles within the frameworks of Einstein-Bardeen and Einstein-Hayward models. We focus particularly on how the physical properties of neutron stars are influenced by the presence of such nonlinear magnetic monopoles. Our results show that under specific conditions and when endowed with sufficient magnetic charge, these systems can exhibit frozen states analogous to those discussed in previous studies \cite{Zhao:2025hdg, Wang:2023tdz, Yue:2023sep, Chicaiza-Medina:2025wul,Brihaye:2025dlq}. This extends the concept of frozen states to compact objects composed of ordinary matter (non-field matter), thereby offering a new perspective for related research. 

The structure of this paper is as follows. In Sec.~\ref{fame}, we introduce the Einstein-Bardeen and Einstein-Hayward NED models with neutron star matter. In Sec.~\ref{nu}, we present numerical solutions and analysis of their physical properties. We conclude and discuss in Sec.~\ref{su}. 
\section{Framework}
\label{fame}
\subsection{The Model}
In this section, we wish to provide a concise introduction to the theoretical framework encompassing the Einstein-nonlinear electrodynamics model, coupled with the matter, described by the following action
\begin{equation} 
\label{act} 
S=\int\sqrt{-g}d^{4}x \left (\frac{c^{3}}{16\pi G}R+\frac{1}{c}\mathcal{L}_{m}+\frac{1}{c}\mathcal{L}_{\text{NED}}\right ),
\end{equation}
where R denotes the scalar curvature, $\mathcal{L}_{m}$ is Lagrangians of the matter, and $\mathcal{L}_{\text{NED}}$ is Lagrangians of the nonlinear electromagnetic field which is 
a function dependent on $F=\frac{1}{4}F_{\mu\nu}F^{\mu\nu}$
involving the electromagnetic field strength $F_{\mu\nu} = \partial_{\mu}A_{\nu}-\partial_{\nu}A_{\mu}$. In this work, we focus on the Einstein-Bardeen and Einstein-Hayward models, whose Lagrangians are respectively
\begin{eqnarray}
\label{Lb} 
\mathcal{L}_{\text{Bardeen}}&=&-\frac{3}{2}\frac{1}{sq^{2}}\left (\frac{\sqrt{2q^{2}FC_{d}}}{1+\sqrt{2q^{2}FC_{d}}}\right )^{\frac{5}{2}}, \\
\label{Lh} 
\mathcal{L}_{\text{Hayward}}&=&-\frac{3}{2}\frac{1}{sq^{2}}\frac{ (2q^{2}FC_{d} )^{\frac{3}{2}}}{(1+(2q^{2}F C_{d})^{\frac{3}{4}})^{2}}. 
\end{eqnarray}
The constants $q$ and $s$ serve as two independent parameters, where $q$ is the magnetic charge. $C_{d}=\frac{(4\pi G)^{2}}{\mu_{0}^{2}c^8}$ is a dimensional constant composed of fundamental physical constants (where $G$ is the gravitational constant, $c$ is the speed of light in vacuum, and $\mu_{0}$ is the vacuum permeability, which ensures that the  $2q^2FC_{d}$ as a whole is dimensionless. When rationalized natural units($4\pi G=c=\mu_{0}=1$) are employed, the result reduces to that of Reference  \cite{Hayward:2005gi, Ayon-Beato:2000mjt}. 

Performing the variation of Eq.~(\ref{act}) with respect to the metric and the electromagnetic field, we obtain the equations of motion:
\begin{eqnarray}
\label{Gx} 
R_{\mu\nu}-\frac{1}{2}g_{\mu\nu}R&=&\frac{8\pi G}{c^{4}}\left (T^{m}_{\mu\nu}+T^{\text{NED}}_{\mu\nu}\right ), \\
 \bigtriangledown_{\mu}\left (\frac{\partial\mathcal{L}_{\text{NED}}}{\partial \mathcal{F}}F_{\mu\nu} \right) &=&0, 
\end{eqnarray}
with
\begin{equation} 
\label{tned}
T^{\text{NED}}_{\mu\nu}=-\frac{\partial\mathcal{L}}{\partial\mathcal{F}}F_{\mu \lambda}F_{\ \nu}^{\lambda}+g_{\mu\nu}\mathcal{L}_{\text{NED}}. 
\end{equation}
% It provides the method for calculating the energy-momentum tensor of nonlinear electromagnetic fields. 

\subsection{Modified Tolman–Oppenheimer–Volkoff Equations}
\label{ssec:nu}
We will consider a static spherically symmetric NS, so we can write the metric as
\begin{equation} 
\label{g}
ds^2=-e^{2\alpha(r)}c^2dt^2+e^{2\beta(r)}dr^2+r^{2}(d\theta^2+\sin^{2}\theta d\varphi ^2). 
\end{equation}

We treat the matter as a perfect fluid with energy-momentum tensor
\begin{equation} 
\label{tm}
T_{\mu\nu}=(\rho c^2+p)U_{\mu}U_{\nu}+pg_{\mu\nu},
\end{equation}
where $\rho c^2$ and $p$ are the energy density and pressure of the matter. And normalized to $U^{\mu}U_{\mu}=-1$, it becomes 
\begin{equation} 
U_{\mu}=(e^{\alpha}, 0, 0, 0).
\end{equation}

Substituting Eq.~(\ref{g}) and Eq.~(\ref{tm}) into Eq.~(\ref{Gx}) yields the following ordinary differential equations from $G_{tt}$, $G_{rr}$ 
% And the prime symbol $'$ denotes the partial derivative with respect to $r$, 
\begin{align}
\frac{ e^{-2 \beta (r)}(2 r \beta '(r)+e^{2 \beta (r)} -1)}{r} &= \frac{8 \pi  G (T^{\text{NED}}_{00}e^{-2 \alpha (r)}+\rho(r) c^2)}{c^4}, \label{G00} \\
\frac{e^{-2 \beta (r)}(2 r \alpha '(r)-e^{2 \beta (r)}+1)}{r^2} &= \frac{8 \pi  G (T^{\text{NED}}_{11} e^{-2 \beta (r)}+p(r)) }{c^4}. \label{G11}
\end{align}

Additionally, we employ the following ansatzes for the nonlinear electromagnetic field, which is solely contributed by nonlinear magnetic monopoles
\begin{equation}
\label{A}
A=q\cos(\theta)d\varphi. 
\end{equation}
Thus, with the ansatz of electromagnetic field in Eq.~(\ref{A}), the magnetic field is given by
\begin{equation} 
\label{F}
F_{\theta\varphi}=-q\sin(\theta). 
\end{equation}

By substituting Eq.~(\ref{Lb}) and Eq.~(\ref{Lh}) into Eq.~(\ref{tned}), followed by substitution into Eq.~(\ref{G00}) under the conventions of Eq.~(\ref{F}), we obtain the corresponding solutions for $e^{-2\beta(r)}$:
\begin{eqnarray}
\label{eq:dedfend} 
e^{-2\beta(r)}_{\text{Bardeen}}=1-\frac{4 \pi  G C_{d} q r^2 \sqrt{\frac{r^2}{\sqrt{C_{d}}}+q^2}}{c^4 s \left(\sqrt{C_{d}} q^2+r^2\right)^2}-\frac{2 G m(r)}{c^2 r},\\
e^{-2\beta(r)}_{\text{Hayward}}=1-\frac{4 \pi  G C_{d}^{\frac{3}{4}} q r^2}{c^4 s \left(C_{d}^{\frac{3}{4}} q^3+r^3\right)}-\frac{2 G m(r)}{c^2 r}.
\end{eqnarray}
When $q=0$, the metric reduces to the Schwarzschild solution. When $m(r)=0$, it recovers the pure Bardeen or Hayward metric. In the following, we take the conventional definition used in neutron star studies  \cite{Tolman:1939jz, Oppenheimer:1939ne}
\begin{equation} 
\label{m} 
m(r)=4\pi \int^{r}_{0}\rho(x)x^2dx. 
\end{equation}
Here, $m(r)$ denotes only the gravitational mass contributed by star matter. 
% Assuming the stellar radius is $R$, the Schwarzschild mass $M$ contributed by ordinary matter is given by 
% \begin{equation} 
% M=m(R)=4\pi \int^{R}_{0}\rho(x)x^2dx. 
% \end{equation}

Ansatzes of the nonlinear electromagnetic field automatically ensure the conservation of the energy-momentum tensor. Consequently, the resulting conservation equation is dictated solely by the matter. The energy-momentum conservation $\bigtriangledown _{\mu}T^{\mu\nu}=\bigtriangledown _{\mu}(T^{\mu\nu}_{m}+T^{\mu\nu}_{\text{NED}})=0$, and it gives
\begin{equation} 
(\rho(r)c^2+p(r))\alpha(r)'+p'(r)=0. 
\end{equation}

By combining this result with Eq.~(\ref{G11}), we eliminate $\alpha'(r)$ to derive the modified Tolman-Oppenheimer-Volkoff  (TOV) equations
\begin{align}
&m'(r) = 4\pi r^2 \rho(r), \label{eq:dedfend1} \\
&\begin{aligned}
& r \left(c^2 \rho(r) + p(r)\right) \cdot \\
&\quad \frac{\left(-\frac{8\pi G\left(T^{\text{NED}}_{11}e^{-2 \beta (r)} + p(r)\right)}{c^4} + \frac{-\frac{2 e^{-2\beta(r)} r p'(r)}{c^2 \rho(r) + p(r)} + e^{-2\beta(r)} - 1}{r^2}\right)}{2 e^{-2\beta(r)}}
\end{aligned} \nonumber \\
&= 0. \label{TOV}
\end{align}

The modified TOV equations are governed by both nonlinear electromagnetic fields and matter. In the limit of vanishing the magnetic charge $q\to 0$, these equations reduce to the original form derived by Tolman, Oppenheimer and Volkoff  \cite{Tolman:1939jz, Oppenheimer:1939ne}.

\section{Numerical Calculation}
\label{nu}

The modified TOV equations will be solved from the center
at $r=0$ to the surface of the star at $r=R$, satisfying
the boundary conditions:
\begin{equation}
m(0) = 0,\  \rho c^2(0) = \rho _c c^2, 
\end{equation}
where $\rho_c$ is the central density. And the neutron star radius $R$ is determined when the pressure vanishes $p(R)=0$. The total
gravitational mass of the compact star matter 
$M$
\begin{equation} 
M\equiv m(R)=4\pi \int^{R}_{0}\rho(x)x^2dx. 
\end{equation}
The total ADM mass, incorporating contributions from the nonlinear magnetic monopoles, can be derived from the $\frac{1}{r}$ term in the asymptotic expansion of the metric at spatial infinity $r\to \infty$. Coincidentally, at spatial infinity $r\to \infty$, Einstein-Bardeen and Einstein-Hayward models exhibit identical asymptotic expansions
\begin{align}
\label{adm}
M_{ADM}=(M+\frac{2\pi C_{d}^{\frac{3}{4}}q }{c^2s}). 
\end{align}

We numerically solved the modified TOV equations using three different equations of state (EOS) to investigate neutron star physical properties under nonlinear magnetic monopoles influence. 
The stiffness of the EoS models employed (BSk19 \cite{Potekhin:2013qqa}, SLy4 \cite{Douchin:2001sv}, AP4 \cite{Akmal:1997ft, Gungor:2011vq}) increases progressively. 
To elucidate the effect of the magnetic charge $q$, we present our numerical results under two different computational schemes:
\begin{enumerate}
\item \textbf{Fixed $sq^2$}: This method follows Ref.~\cite{Zhao:2025hdg, Wang:2023tdz, Yue:2023sep} and has been demonstrated to be  straightforward, effective and  reliable.
\item \textbf{Fixed $s$}: This method, which fixes the coupling parameter, is more physically fundamental but restricts the $q$ parameter space.

\end{enumerate}

During numerical solution, we find that for a fixed value of either  $sq^2$ or $s$, there exists the critical magnetic charge $q_{c}$. Upon reaching the critical magnetic charge $q_{c}$, neutron stars transitions into frozen states. Beyond this critical magnetic charge $q_{c}$, it becomes numerically infeasible to obtain physically meaningful solutions. The magnitude of this critical magnetic charge depends on the nonlinear electromagnetic field model, the equation of state, and the central density, as detailed in Tab.~\ref{bqc} and Tab.~\ref{hqc}. Clearly, the softer the equation of state and the higher the central density, the smaller the critical magnetic charge. 
\begin{table}[htbp] 
\renewcommand\arraystretch{1.5}
\captionsetup{justification=raggedright}
\caption{The critical magnetic charge in the Bardeen framework for fixed $s=4. 75*10^{-81}[\rm{m^3J^{-1}Wb^{-2}}]$ or fixed $sq^2=4. 75*10^{-81}[\rm{m^3J^{-1}Wb^{-2}}]q_c^2$, for various equations of state and central densities. }
\label{bqc}
\begin{tabular}{ |c | c  c  c  | } 
\hline 
$\rho_{c}$ & $0. 5\times 10^{18}$ & $1. 0\times 10^{18}$ & $1. 5\times 10^{18}\ [\rm{kg/m^{3}}]$ \\ 
\hline 
BSk19& $2. 848\times 10^{22}$ & $2. 5685\times 10^{22}$ & $2. 3310\times 10^{22}\ [\rm{Wb}]$ \\
 SLy4& $3. 0774\times 10^{22}$ & $2. 7161\times 10^{22}$ & $2. 4322\times 10^{22}\ [\rm{Wb}]$ \\
AP4 & $3. 1608\times 10^{22}$ & $2. 8335\times 10^{22}$ & $\ \ \ \ -- \ \ \ \ \ \ \ \ \  [\rm{Wb}]$ \\
\hline 

\end{tabular}
\end{table}
\begin{table}[htbp] 
\renewcommand\arraystretch{1.5}
\captionsetup{justification=raggedright}
\caption{The critical magnetic charge in the Hayward framework for fixed $s=7*10^{-81}[\rm{m^3J^{-1}Wb^{-2}}]$ or fixed $sq^2=7*10^{-81}[\rm{m^3J^{-1}Wb^{-2}}]q_c^2$, for various equations of state and central densities. }
\label{hqc}
\begin{tabular}{ |c | c  c  c  | } 
\hline 
$\rho_{c}$ & $0. 5\times 10^{18}$ & $1. 0\times 10^{18}$ & $1. 5\times 10^{18}\ [\rm{kg/m^{3}}]$ \\ 
\hline 
BSk19& $3. 0774\times 10^{22}$ & $2. 6440\times 10^{22}$ & $2. 4911\times 10^{22}\ [\rm{Wb}]$ \\
 SLy4& $3. 1801\times 10^{22}$ & $2. 7161\times 10^{22}$ & $2. 4322\times 10^{22}\ [\rm{Wb}]$ \\
AP4 & $3. 2556\times 10^{22}$ & $2. 9022\times 10^{22}$ & $\ \ \ \ -- \ \ \ \ \ \ \ \ \  [\rm{Wb}]$ \\
\hline 

\end{tabular}
\end{table}

In following subsections, we will demonstrate that reaching the critical magnetic charge indeed results in frozen states, and we further analyze its physical properties.

\subsection{Radial Pressure Profiles} 
\label{RPP}
The radial pressure distribution was obtained directly by solving the modified Tolman-Oppenheimer-Volkoff (TOV) equation.

\begin{figure*}[]
\begin{center}

\subfloat{\includegraphics[width=0.85\textwidth]{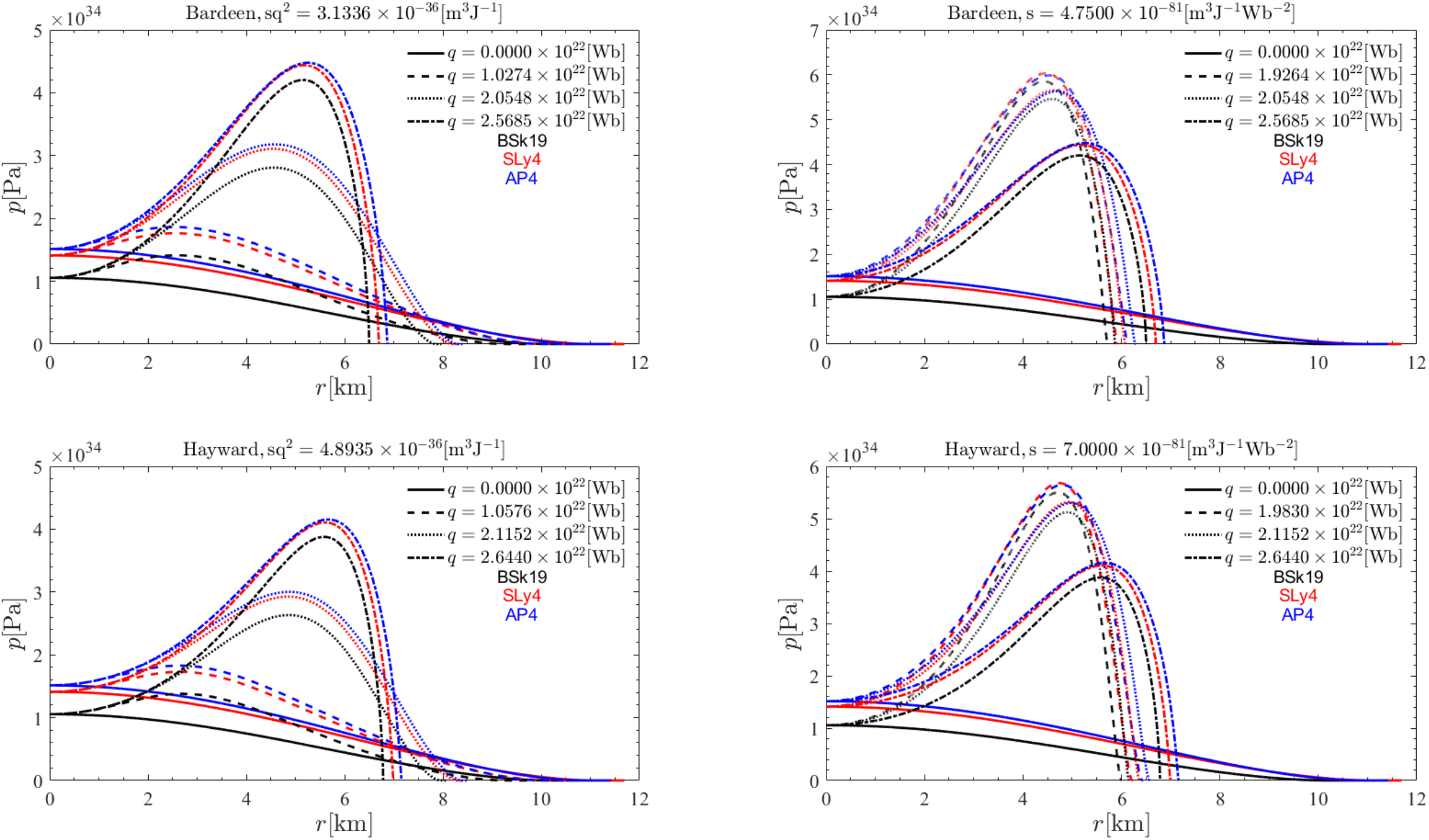}}
\end{center}
\captionsetup{justification=raggedright}
\caption{The top subplots show the radial pressure profiles in the Einstein-Bardeen framework with $\rho_{c}=1. 0\times10^{18}[\rm{kg / m^{3}}]$: (left) as a function of $q$ with fixed $sq^2=3. 1336\times10^{-36}[\rm{m^3J^{-1}}]$ , (right) as a function of $q$ with fixed $s=4. 7500\times10^{-81}[\rm{m^3J^{-1}Wb^{-2}}]$ . The bottom subplots present corresponding results for the Einstein-Hayward framework with fixed $sq^2=4. 8935\times10^{-36}[\rm{m^3J^{-1}}]$ (left)  and fixed $s=7. 0000\times10^{-81}[\rm{m^3J^{-1}Wb^{-2}}]$ (right). }
\label{pr}
\end{figure*}

Fig.~\ref{pr} presents the radial pressure profiles in neutron star matter after incorporating nonlinear magnetic monopoles in $\rho_{c}=1. 0\times10^{18}[\rm{kg/m^{3}}]$. For both the Einstein-Bardeen and Einstein-Hayward frameworks with identical fixed parameters, the variation of the radial pressure profile with magnetic charge $
q$ is consistent.
 
Observations from Fig.~\ref{pr}  (left subplots) under fixed $s q^2$ clearly demonstrate two key effects of increasing the magnetic charge $q$: 
\begin{enumerate}
\item \textbf{Radial contraction}: The neutron star boundary exhibits definitive inward contraction. 
\item \textbf{Pressure enhancement}: A progressively pronounced high-pressure region develops near the boundary. 
\end{enumerate}
 This phenomenon is attributed to the negative pressure and the additional gravitational potential supplied by the nonlinear magnetic monopoles as Eq.~(\ref{TOV}). Because of the positive correlation between density and pressure in the equation of state, Fig.~\ref{pr} suggests that as the magnetic charge $q$ gradually increases and approaches the critical magnetic charge, a dense outer layer forms, resulting in ``filled hard candy like" structure. 

 % For the fixed coupling parameter $s$, Fig.~\ref{pr} (right subplots) show that as the magnetic charge $q$ increases, the maximum central pressure decreases while the stellar radius increases—behavior strikingly different from the fixed $sq^2$ case. This occurs because, with fixed $s$, the term $\frac{1}{sq^2}$ in Eq.~(\ref{Lb}) and Eq.~(\ref{Lh}) decreases with increasing $q$, suppressing the contribution of the nonlinear electromagnetic field, while the remaining terms show relative little variation in this regime. As the magnetic charge $q$ decreases, the field’s influence strengthens, raising the maximum pressure.
 For the fixed coupling parameter $s$, Fig.~\ref{pr} (right subplots) shows that as the magnetic charge $q$ decreases, the maximum central pressure increases while the stellar radius decreases—behavior strikingly different from the fixed $sq^2$ case. This occurs because, with fixed $s$, the term $\frac{1}{sq^2}$ in Eq.~(\ref{Lb}) and Eq.~(\ref{Lh}) increases with decreasing $q$, enhancing the contribution of the nonlinear electromagnetic field, while the remaining terms show relatively little variation in this regime. However, the causality constraint on the equation of state imposes an upper bound on the pressure, thereby restricting the parameter space of $q$ that yields physically meaningful solutions and preventing its arbitrary reduction. Especially causality constraint limits the maximum pressure attainable by a stiffer equation of state (AP4) strongly, resulting in a lower value than other EOSs (Tab.~\ref{eosp}). Consequently, at a central density of $1. 5\times 10^{18}[\rm{kg/m^3}]$, maximum allowed pressure of AP4 cannot support frozen states, and its critical magnetic charge is therefore absent from Tab.~\ref{bqc} and Tab.~\ref{hqc}.
 
 % As the magnetic charge $q$ decreases, the field’s influence strengthens, raising the maximum pressure in radial pressure profiles. 

\begin{table}[htbp] 
\renewcommand\arraystretch{1.5}
\captionsetup{justification=raggedright}
\caption{Maximum allowed pressure from causality-constrained equations of state.}
\label{eosp}
\begin{tabular}{ |c | c  c  c  | } 
\hline 
EOS & BSk19 & SLy4 & AP4 \\ 
\hline 
$p_{max}$& $1.6149\times 10^{35}$ & $1. 5107\times 10^{35}$ & $6.3972\times10^{34}\ [\rm{Pa}]$ \\
\hline 
\end{tabular}
\end{table}

\subsection{Compactness $\mathcal{C}$/Average Density $\hat{\rho}-q$   Relationship} 
\begin{figure*}[]
\begin{center}
\subfloat{\includegraphics[width=0.85\textwidth]{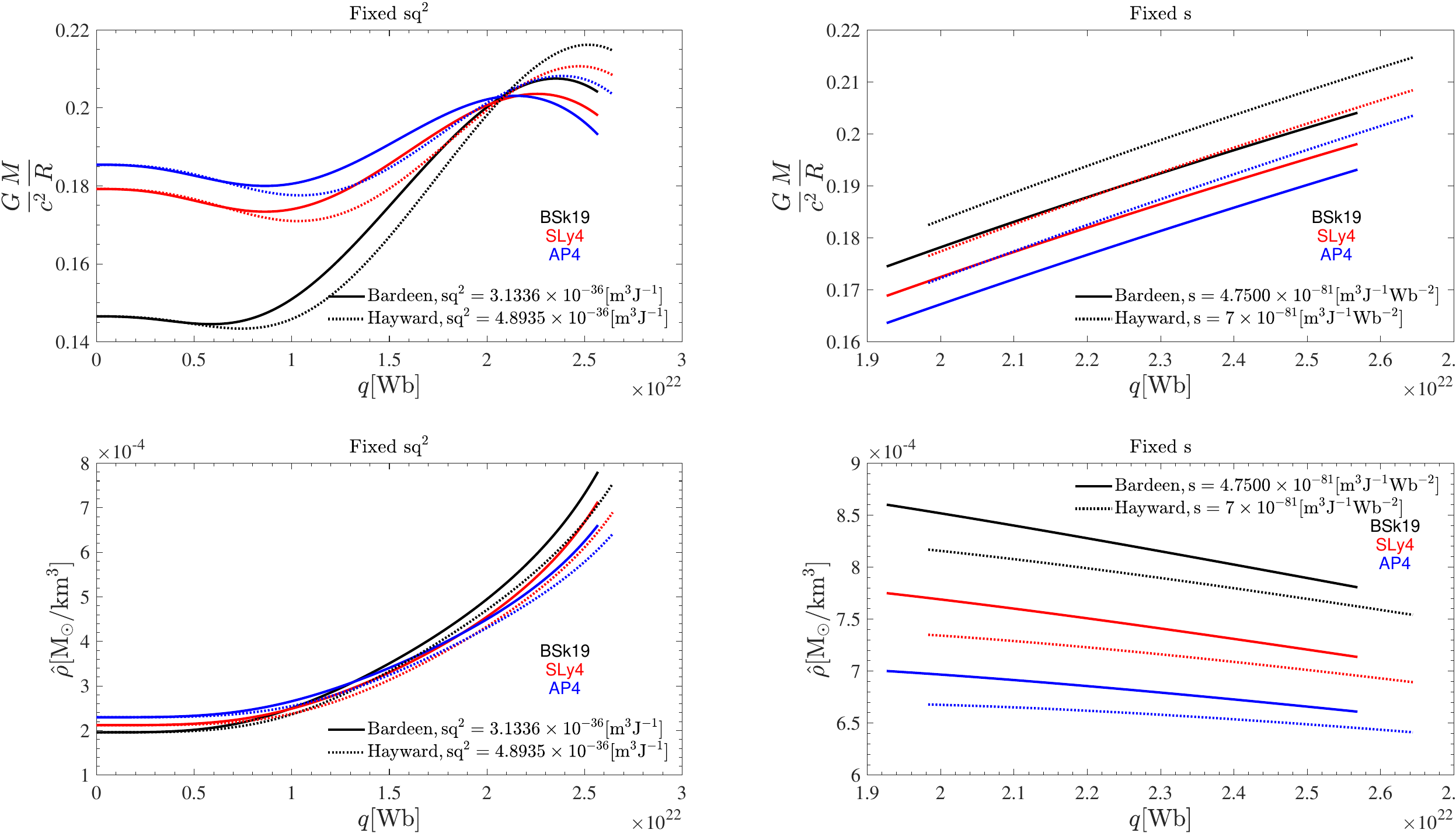}}
\end{center}
\captionsetup{justification=raggedright}
\caption{ The top upper subplots display the variation of the compactness $\mathcal{C}$ with magnetic charge $q$ at central density $\rho_{c}=1. 0\times10^{18}[\rm{kg / m^{3}}]$, under the fixed $sq^2$ and fixed $s$ schemes. The two bottom subplots display the variation of the average density $\hat{\rho}$ with magnetic charge $q$ at central density $\rho_{c}=1. 0\times10^{18}[\rm{kg / m^{3}}]$, under the fixed $sq^2$ and fixed $s$ schemes. }
\label{cr}
\end{figure*}

In this section, we discuss the relationship between compactness $\mathcal{C}$, average density $\hat{\rho}$, and the magnetic charge $q$. Fig.~\ref{cr} present the variations of compactness and average density with $q$ in the Einstein-Bardeen and Einstein-Hayward frameworks, respectively, under the condition $\rho_c = 1. 0\times 10^{18}[\rm{kg/m^3}]$. 

As illustrated in the corresponding Fig.~\ref{cr}, the average density $\hat{\rho}$ demonstrates a distinct dual behavior under different constraints: it increases with $q$ when $s q^2$ is held fixed, but decreases with $q$ when only $s$ is fixed. Due to the positive correlation between pressure and density in the equation of state, this result was already implied in Fig.~\ref{pr}.

Both frameworks exhibit consistent trends in how compactness $\mathcal{C}$ varies with magnetic charge $q$: when $sq^2$ is fixed, the compactness first decreases slightly, then increases substantially, and finally undergoes another slight decrease as $q$ increases. The amplitude of these variations shows a clear dependence on the stiffness of the equation of state: the softer BSk19 model exhibits the smallest initial decrease and the most dramatic subsequent increase, whereas the stiffer AP4 model shows the largest initial decrease and the least pronounced growth. 

This nonmonotonic behavior results from a competition between the neutron star radius $R$ and the density $\hat{\rho}$, driven by the nonlinear magnetic monopoles. A straightforward analysis can be conducted using the following relation:
\begin{align}
\label{c}
\mathcal{C}=\frac{G M}{c^2 R}= \frac{G}{c^2 }\frac{M}{4/3\pi R^3}\frac{4/3\pi R^3}{R}\propto\hat{\rho}R^2. 
\end{align}
The variation of compactness $\mathcal{C}$ is governed by distinct mechanisms in different regimes of $q$. 
At small $q$, the decrease in compactness is dominated by the reduction in $R$ caused by the nonlinear  magnetic monopoles. 
 At intermediate $q$, the increase in compactness is primarily driven by the growth in average density $\hat{\rho}$. 
 Near the critical magnetic charge $q_c$, the reduction in $R$ again becomes the dominant factor, leading to a slight decrease in compactness.

Under fixed $s$, the compactness $\mathcal{C}$ increases monotonically with the magnetic charge $q$ within its allowed range. This behavior arises because the reduction in the neutron star radius $R$ dominates the variation of $C$, despite a concomitant decrease in the average density $\hat{\rho}$ with increasing $q$. 
% Eq.~(\ref{c}) implies that the $q$-induced increase of $R$ outweighs the effect of the change in $\hat{\rho}$, leading to a net increase in compactness. 

\subsection{Radial Metric Profiles} 
\label{state}
\begin{figure*}[]
\begin{center}
\label{fig:grr}
\subfloat{\includegraphics[width=0.85\textwidth]{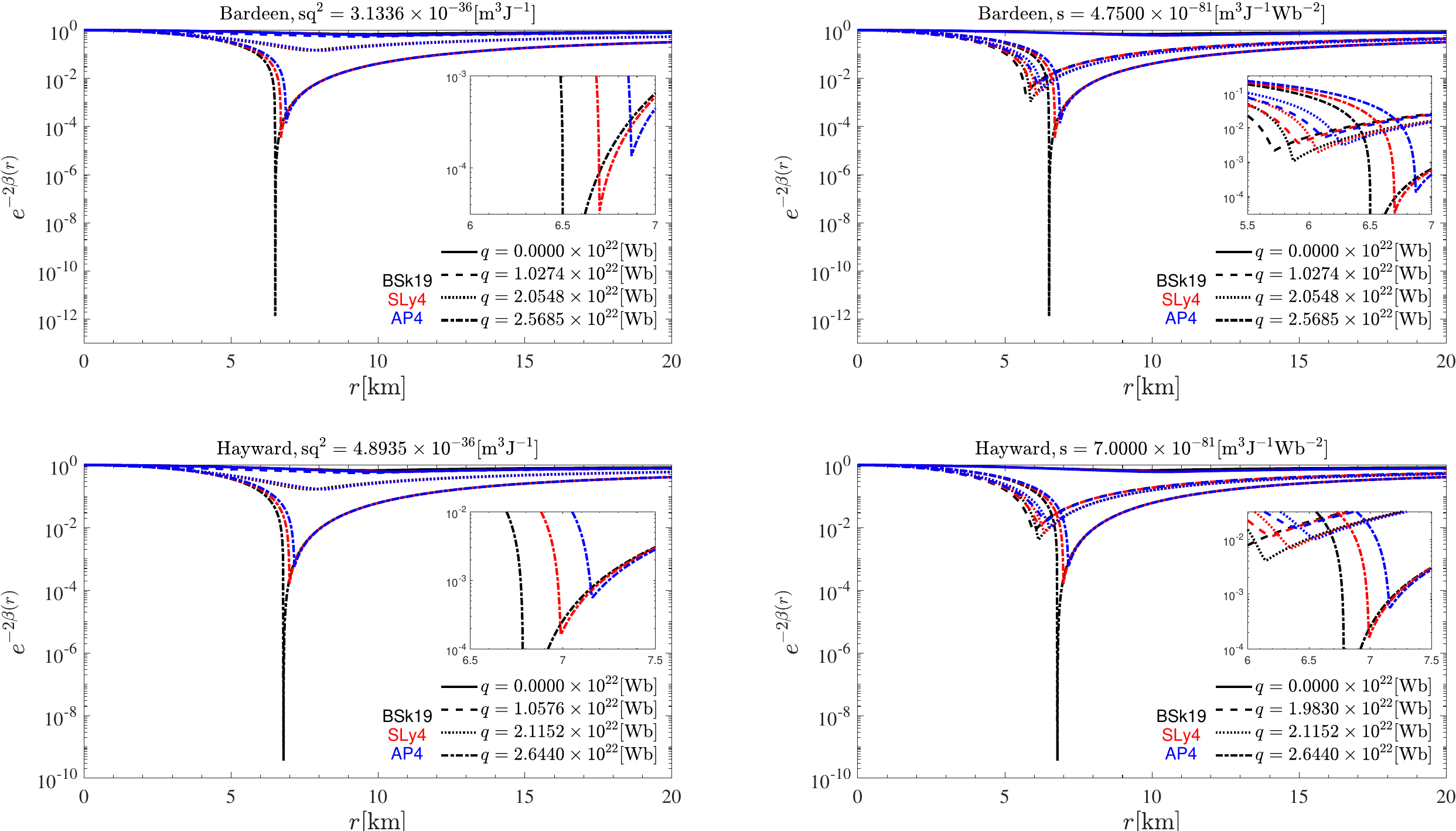}}
\end{center}
\captionsetup{justification=raggedright}
\caption{The top subplots show $e^{-2\beta(r)}$($1/g_{rr}$) in the Einstein-Bardeen framework with $\rho_{c}=1. 0\times10^{18}[\rm{kg / m^{3}}]$: (left) as a function of $q$ with fixed $sq^2=3. 1336\times10^{-36}[\rm{m^3J^{-1}}]$ , (right) as a function of $q$ with fixed $s=4. 7500\times10^{-81}[\rm{m^3J^{-1}Wb^{-2}}]$ . The bottom subplots present corresponding results for the Einstein-Hayward framework with fixed $sq^2=4. 8935\times10^{-36}[\rm{m^3J^{-1}}]$ (left)  and fixed $s=7. 0000\times10^{-81}[\rm{m^3J^{-1}Wb^{-2}}]$ (right). }
\label{grr}
\end{figure*}

\begin{figure*}[]
\begin{center}

\subfloat{\includegraphics[width=0.85\textwidth]{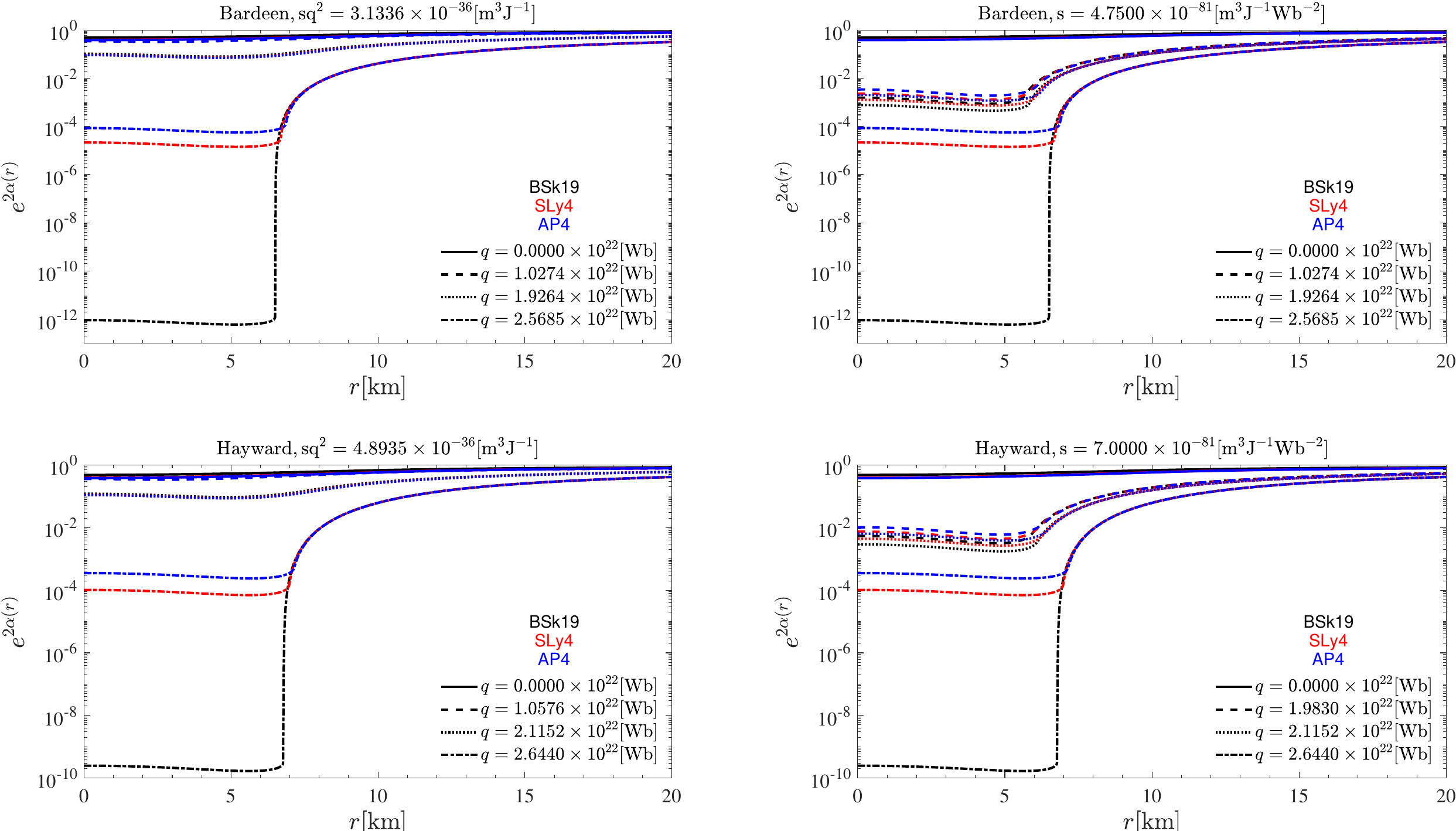}}
\end{center}
\captionsetup{justification=raggedright}
\caption{The top subplots show $e^{2\alpha(r)}$($-g_{tt}$) in the Einstein-Bardeen framework with $\rho_{c}=1. 0\times10^{18}[\rm{kg / m^{3}}]$: (left) as a function of $q$ with fixed $sq^2=3. 1336\times10^{-36}[\rm{m^3J^{-1}}]$ , (right) as a function of $q$ with fixed $s=4. 7500\times10^{-81}[\rm{m^3J^{-1}Wb^{-2}}]$ . The bottom subplots present corresponding results for the Einstein-Hayward framework with fixed $sq^2=4. 8935\times10^{-36}[\rm{m^3J^{-1}}]$ (left)  and fixed $s=7. 0000\times10^{-81}[\rm{m^3J^{-1}Wb^{-2}}]$ (right). }
\label{gtt}
\end{figure*}

In this section, we examine the radial profile of the metric. Fig.~\ref{gtt} and Fig.~\ref{grr} display the behavior of the metric functions $e^{2\alpha(r)}$ ($-g_{tt}$) and $e^{-2\beta(r)}$ ($1/g_{rr}$) with central density $\rho_c=1. 0\times 10^{18}[\rm{kg/m^3}]$, respectively. 

For both the Einstein-Bardeen and Einstein-Hayward frameworks, the variation of metric functions with magnetic charge $q$ is consistent. When $s q^2$ is fixed, the position of the minimum $1/g_{rr}$ gradually shifts inward as the magnetic charge $q$ increases, which is consistent with the trend of neutron star radius changes as shown in Fig.~\ref{pr}, and the minimum value decreases progressively. Similarly, when $s$ is fixed, the position of the minimum $1/g_{rr}$ moves outward with increasing magnetic charge $q$, again aligning with the trend of neutron star radius variations as shown in Fig.~\ref{pr}, while the minimum value also decreases. 

For BSk19, the critical magnetic charge $q_{c}$ is the smallest among the three equations of state. We ttherefore show the metric solutions at its critical magnetic charge $q_{c}$, $q=2. 5685\times 10^{22}[\rm{Wb}]$ for Einstein-Bardeen framework and $q=2. 6440\times 10^{22}[\rm{Wb}]$ for Einstein-Hayward framework in Fig.~\ref{grr} and Fig.~\ref{gtt}.For BSk19 at the critical magnetic charge $q_{c}$, the minimum value of $1/g_{rr}$ is observed to be extremely close to zero, reaching magnitudes as low as $10^{-9}$ to $10^{-12}$ for both models. At the critical magnetic charge $q_{c}$, the location of this $1/g_{rr}$ minimum coincides with the neutron star boundary, indicating that all matter is confined within this radius. Moreover, inside this radius, $-g_{tt}$ also approaches zero extremely closely, signifying that the neutron star has entered frozen states. In summary, the position of this $1/g_{rr}$ minimum exhibits properties similar to those of the critical horizon (as discussed in Ref.~\cite{Zhao:2025hdg, Wang:2023tdz, Yue:2023sep}), and can be identified as the critical horizon of frozen neutron stars.

\subsection{Mass-Radius Relation}
\begin{figure*}[]
\begin{center}

\subfloat{\includegraphics[width=0.85\textwidth]{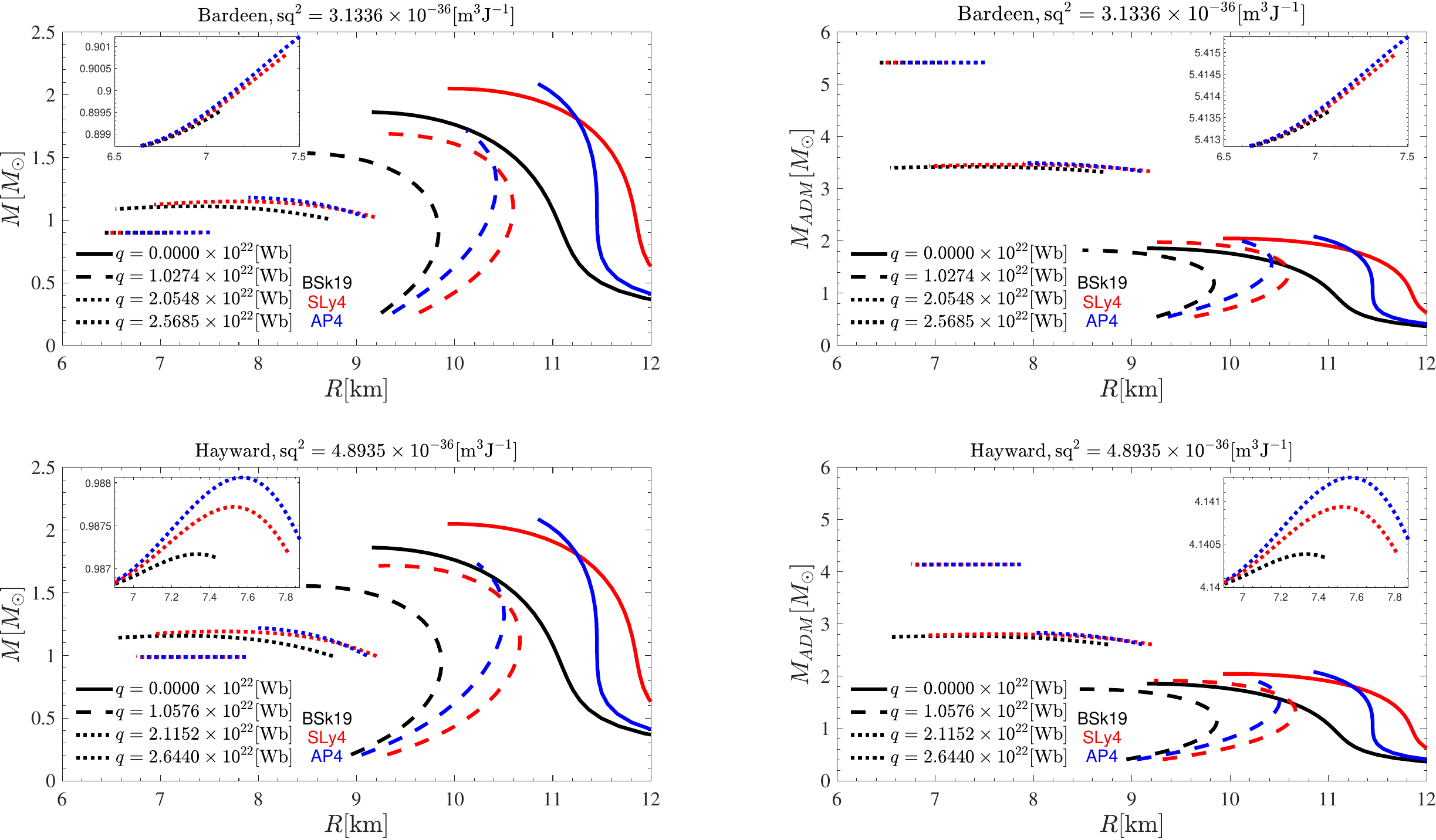}}
\end{center}
\captionsetup{justification=raggedright}
\caption{The left subplots shows the dependence of the $M$-$R$ relation on the magnetic charge $q$ under the fixed $sq^2$ scheme, while the right subplot displays the corresponding variation of the $M_{ADM}$-$R$ relation with $q$ under the same scheme. The top subplots corresponds to the Einstein-Bardeen framework, while the bottom subplots corresponds to the Einstein-Hayward framework. }
\label{mrsq2}
\end{figure*}

The mass-radius ($M$-$R$) relation is a crucial issue in neutron star research and serves as the most direct relation for constraining theories. In the left subplots of Fig.~\ref{mrsq2}, we show the modification of the $M$-$R$ relation for different magnetic charges $q$ at fixed $sq^2$, while the right subplots presents the corresponding modification for the $M_{ADM}$-$R$ relation. As the magnetic charge increases, the $M$-$R$ relation exhibits significant deformation: the allowed radius of neutron star solutions gradually decrease, and the range of possible $M$ values also narrows. A similar trend is observed for the $M_{ADM}$-$R$ relation, but with a key distinction: as the magnetic charge $q$ increases, the maximum value of $M$ gradually decreases, primarily due to the reduction in neutron star radius $R   $ dominating the decrease in maximum mass. In contrast, the maximum $M_{ADM}$ increases steadily, owing to the growing contribution of the magnetic charge $q$ to $M_{ADM}$, as indicated by Eq.~(\ref{adm}). Indeed, when the neutron star reaches frozen states, the primary contribution to its $M_{ADM}$ originates from the magnetic charge $q$, as showed in Tab.~\ref{madm}. 
\begin{table}[htbp] 
\renewcommand\arraystretch{1.5}
\captionsetup{justification=raggedright}
\caption{When BSk19 reaches the critical magnetic charge $q_{c}$ and forms the frozen neutron star, the composition of $M_{ADM}$ is as summarized (under parameter conditions consistent with Sec.~\ref{state}). }
\label{madm}
\begin{tabular}{ |c | c  c  c  | } 
\hline 
 & $M$ & $\frac{2\pi C_{d}^{\frac{3}{4}}q }{c^2s}$ & $M_{ADM}$ \\ 
\hline 
$Bardeen$& $0. 8995$ & $4. 5141$ & $5. 4136\ [\rm{M_{\odot}}]$ \\
$Hayward$& $0. 9868$ & $3. 1532$ & $4. 1400\ [\rm{M_{\odot}}]$ \\
\hline 
\end{tabular}
\end{table}

\begin{figure*}[]
\begin{center}

\label{sec:SD}
\subfloat{\includegraphics[width=0.85\textwidth]{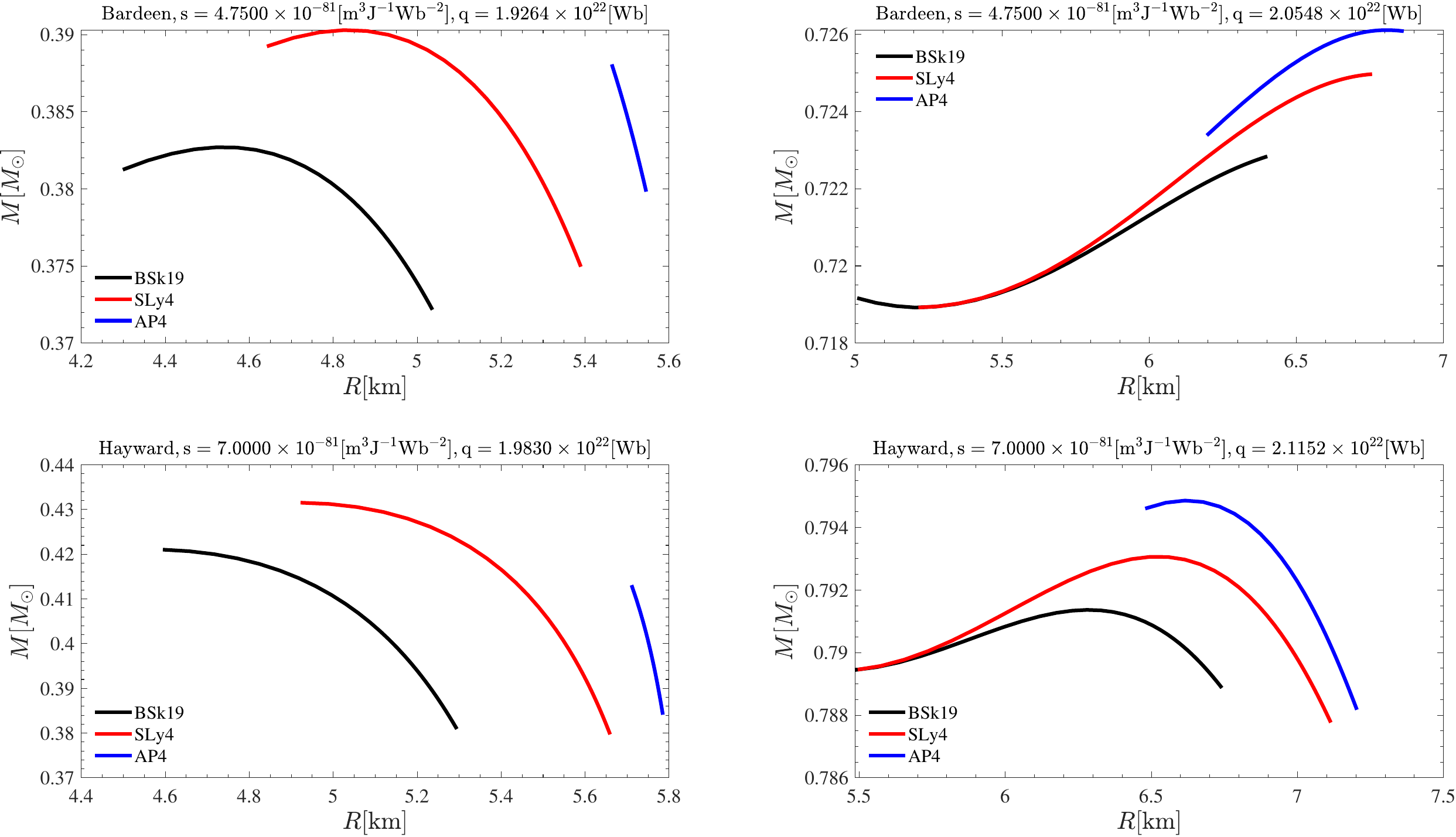}}
\end{center}
\captionsetup{justification=raggedright}
\caption{The $M$-$R$ relation under different values of magnetic charge $q$ with fixed $s$. 
The top subplots corresponds to the Einstein-Bardeen framework, while the bottom subplots corresponds to the Einstein-Hayward framework. }
\label{mrs}
\end{figure*}

Fig.~\ref{mrs} illustrates the variation in the $M$–$R$ relation for a fixed value of $s$. Within the allowed parameter range, the maximum mass $M$ increases with growing $q$ ($M_{ADM}$ also increases). Furthermore, due to the causality constraint discussed in Sec.~\ref{RPP}, the allowed ranges of $M$ and $R$ for the AP4 model are significantly reduced at small values of $q$.

Although the inclusion of nonlinear magnetic monopoles leads to significant deviations in the neutron star $M–R$ relation and the $M_{ADM}–R$ relation from those predicted by pure general relativity without nonlinear magnetic monopoles, the additional degree of freedom introduced by the magnetic charge $q$, together with the distinctive properties of frozen neutron stars, complicates the identification and classification of such objects. 

\section{Summary and Discussion}
\label{su}
In this work, we have systematically investigated the structural and gravitational properties of neutron stars incorporating nonlinear magnetic monopoles within the frameworks of Einstein-Bardeen and Einstein-Hayward nonlinear electrodynamics. By solving the modified Tolman-Oppenheimer-Volkoff equations under both fixed coupling parameter $s$ and fixed $sq^2$ schemes, we find that the presence of a magnetic charge $q$ significantly deforms, radial pressure profiles, spacetime geometry of neutron stars and the mass-radius relation. Under fixed $sq^2$, increasing $q$ leads to substantial radial contraction and the formation of a high-density surface layer, yielding more compact configurations. Under fixed $s$, the neutron stars cexpands while average density decreases and compactness $\mathcal{C}$ increases with increasing $q$. The inclusion of nonlinear magnetic
monopoles leads to significant deviations in the neutron
star $M –R$ relation and the $M_{ADM}–R$ relation from
those predicted by pure general relativity without nonlinear magnetic monopoles. 

Most notably, beyond the critical magnetic charge $q_{c}$, neutron stars transition into frozen states characterized by the formation of the critical horizon—a configuration in which both metric components vanish extremely close to zero at the neutron stars radius: specifically, $1/g_{rr}$ becomes nearly zero at the surface, while $-g_{tt}$ approaches zeros throughout the interior and up to the surface. In conclusion, this work generalizes the concept of frozen states from boson stars to ordinary matter systems, demonstrating that nonlinear magnetic monopoles can modify neutron star structure. 

Several extensions of our study warrant exploration. Firstly, a rigorous analysis of the stability of these frozen neutron stars under radial perturbations is essential. Secondly, the astrophysical implications of such objects—including their formation channels, observable electromagnetic signatures, and potential as gravitational wave sources—remain to be thoroughly investigated. Finally, modified theories of gravity may also give rise to frozen states through mechanisms analogous to nonlinear magnetic monopoles \cite{Wang:2024ehd, Ma:2024olw}. Those intriguing possibility will be addressed in our future work.

\begin{acknowledgments}
This work is supported by the National Natural Science Foundation of China (Grant No. 12275110 and No. 12247101) and the National Key Research and Development Program of China (Grant No. 2022YFC2204101 and 2020YFC2201503). 
\end{acknowledgments}

\end{document}